\documentclass[pra,twocolumn,superscriptaddress,floatfix,longbibliography]{revtex4-1}

\usepackage[utf8]{inputenc}
\usepackage[T1]{fontenc}
\usepackage{graphicx, placeins}
\usepackage{color}
\usepackage{comment}
\usepackage{amsmath}
\usepackage{braket}

\begin{document}

\title{Fast collisional $\sqrt{\mathrm{SWAP}}$ gate for fermionic atoms in an optical superlattice}

\author{Rafi Weill} 
\author{Jonathan Nemirovsky} 
\author{Yoav Sagi}

\email[Electronic address: ]{yoavsagi@technion.ac.il}

\affiliation{Physics Department and Solid State Institute, Technion - Israel Institute of Technology, Haifa 32000, Israel}

\date{\today}
\begin{abstract}
Collisional gates in optical superlattices have recently achieved record fidelities, but their operation times are typically limited by tunneling. Here we propose and analyze an alternative route to a fast $\sqrt{\mathrm{SWAP}}$ gate for two fermionic atoms in an optical superlattice based on optimized, time-dependent control of the short and long lattice depths. The gate is implemented by transiently releasing the atoms into a quasi-harmonic confinement centered between the two sites. With an appropriately chosen contact interaction strength, a controlled collision accumulates the exchange phase required for $\sqrt{\mathrm{SWAP}}$ and generates entanglement. We employ a continuum, time-dependent Schr\"odinger-equation simulation that goes beyond a two-site Fermi--Hubbard description and benchmark it against experimentally implemented tunneling-based protocols, reproducing the observed single-particle tunneling and spin-exchange dynamics. For experimentally accessible lattice depths, we find that the proposed gate operates in $\sim 21\,\mu\mathrm{s}$, more than an order of magnitude faster than tunneling-based implementations, while achieving fidelities $\gtrsim 99\%$. We further analyze sensitivity to lattice-depth variations and show that a composite sequence improves robustness. Our results establish fast, collision-mediated entangling gates in superlattices as a promising building block for scalable neutral-atom quantum computation.
\end{abstract}

\maketitle

\section{Introduction}

Numerical methods that harness the intrinsic quantum properties of physical systems are expected to offer advantages over classical computation in a range of applications, including optimization \cite{lucas2014ising,egger2021warm,buonaiuto2023portfolio}, cryptography \cite{shor1997polynomial,bernstein2009pqc}, complex material simulations \cite{aspuru2005simulated,mcardle2020quantumchem,clinton2024nearterm}, and machine learning \cite{biamonte2017qml,melnikov2023qmlsoftware,peralgarcia2024qmlslr}. In particular, quantum simulations of electronic structures and strongly correlated fermionic phases represent one of the most promising frontiers for demonstrating quantum advantage \cite{bravyi2002fermionic, gonzalez2023fermionic, babbush2023quantumdynamics}. While such fermionic models can, in principle, be simulated on a universal quantum computer, the required overhead is substantial \cite{bravyi2002fermionic,Seeley2012}. An alternative approach is offered by fermionic quantum processors, which provide a more natural framework by directly encoding antisymmetry and particle statistics.

Neutral atoms confined in optical potentials have established themselves as a prominent hardware platform for quantum simulation and computation \cite{Bloch2008}. Two main architectures are currently being pursued: movable tweezer arrays, which enable flexible, reconfigurable geometries \cite{bluvstein2022quantum,manetsch2024tweezer,kotochigova2025neutral,gonzalez2023fermionic,endres2021quantum,pause2024supercharged}, and optical lattices, which provide large-scale, highly ordered registers \cite{park2022cavity,tao2024highfidelity,norcia2024iterative,gyger2024continuous, Bojovic2026}. Their key advantages include long coherence times, inherent scalability, and the ability to realize strong, tunable interactions. Substantial progress has been achieved through the use of Rydberg interactions, enabling fast and high-fidelity two-qubit gate operations \cite{wilk2010entanglement, isenhower2010demonstration, radnaev2025universal,finkelstein2024universal}. In this approach, each qubit is localized in an individual trap, with gate interactions mediated by long-range dipole–dipole coupling. These gates operate on microsecond timescales and are relatively insensitive to the thermal motion of the atoms. Nonetheless, they rely on complex optical excitation schemes to access the Rydberg manifold, and their fidelities are ultimately limited by spontaneous decay from Rydberg or intermediate states.

An alternative approach, known as collisional gates, achieves entanglement in spin and orbital degrees of freedom through controlled atomic motion and contact interactions \cite{mandel2003controlled, folling2007secondorder, anderlini2007exchange,  trotzky2008superexchange, zhang2023spinexchange, Nemirovsky}. Recent experiments have demonstrated collisional entangling gates with fidelities approaching 99.75\% and Bell-state lifetimes exceeding 10 seconds, achieved through coherent manipulation of fermionic atoms in an optical superlattice \cite{Bojovic2026}. A central mechanism is the composite pair-exchange operation, which relies on the exchange blockade: the symmetry of the two-particle wave function, together with on-site interactions, gives rise to spin entanglement when tunneling is enabled between neighboring sites. The gate duration is primarily determined by the tunneling rate, which depends on the inter-site barrier height controlled by the lattice depth. Adjusting the barrier allows tunneling to be effectively switched on and off. Achieving high fidelity requires minimizing motional excitations during these ramps. While strictly adiabatic changes suppress such excitations, they lead to slower gates and limit circuit-level performance \cite{hayes2007quantum}. In practice, shaped pulses of quasi-adiabatic nature have been shown to provide a compromise between speed and fidelity \cite{Bojovic2026}.

Here, we present an alternative route to implementing a fast and robust 
$\sqrt{\mathrm{SWAP}}$ gate for atoms interacting via contact-like potentials 
in an optical superlattice. The concept builds on our previous work in a 
tweezer-based architecture~\cite{Nemirovsky}, where the strategy was to switch 
off the individual tweezers while simultaneously turning on an auxiliary 
harmonic well centered between them. In this configuration, the atomic wave 
packets evolve as squeezed coherent states and are subsequently recaptured 
when the tweezers are restored. The key advantage of that approach is that the 
gate timescale is determined by the harmonic confinement, allowing operations 
much faster than those limited by tunneling. In a lattice-based setting, 
however, one cannot directly engineer a perfectly harmonic well; the potential 
must remain a superlattice \cite{chalopin2025optical} at all times,
\begin{equation}\label{eq:super}
V(x,\tau)= V_S(\tau)\,\cos^2\!\left(\frac{\pi x}{a_x}\right)
+ V_L(\tau)\,\sin^2\!\left(\frac{\pi x}{2a_x}\right),
\end{equation}
where $V_S$ and $V_L$ denote the short- and long-lattice depths, respectively, 
and $a_x$ is the short-lattice constant.

A straightforward extension of our previous approach would be to completely 
turn off the short lattice, allow the atoms to evolve for approximately a 
single transit time, and then restore it. This strategy, however, proves 
ineffective: the residual long-lattice potential is strongly anharmonic, 
causing the wave function to disperse into higher bands and leading to a rapid 
loss of fidelity. In contrast, we demonstrate that retaining a modest 
short-lattice amplitude yields a quasi-harmonic confinement, and that carefully 
engineered time-dependent profiles of $V_S(\tau)$ and $V_L(\tau)$ can support 
high-fidelity dynamics while still allowing for fast gate operation.

The remainder of this paper is organized as follows. In Sec.~\ref{sec_review}, we review the analytical intuition developed in our previous tweezer-based protocol and the underlying harmonic dynamics in a single well. In Sec.~\ref{sec_bloch}, we benchmark and validate our continuum numerical simulations against the experimental results for tunneling-based gate sequences reported in Ref.~\cite{Bojovic2026}. In Sec.~\ref{sec_swap}, we demonstrate a near-ideal SWAP gate in a superlattice using optimized control of $(V_S,V_L)$, thereby laying the groundwork for the main result of this work: a fast two-qubit $\sqrt{\mathrm{SWAP}}$ gate in a superlattice, presented in Sec.~\ref{sec_sqrt_swap}. The robustness of the gate to small variations in the lattice amplitudes and its fidelity within the computational subspace are discussed in Secs.~\ref{sec:Robustness to lattice-amplitude variations} and \ref{sec:Gate Performance in the Computational Subspace}, respectively. In Sec.~\ref{sec:Repeated Gate Performance}, we investigate repeated application of the gate and demonstrate the sensitivity of the resulting fidelity to the waiting times between successive gate operations. Finally, in Sec.~\ref{sec_discussion}, we conclude with a discussion of the implications and outlook.

\section{Fast collisional gate in optical tweezers}\label{sec_review}

We first give a short summary of our previous work \cite{Nemirovsky}, where we proposed a two-atom gate protocol based on switching two micro-traps into a central harmonic trap. After half a trap period, the system is restored to the initial trap configuration. When the two atoms are non-interacting, this sequence acts as a SWAP gate. With tunable contact interactions, the colliding atoms accumulate a relative $\pi/2$ phase shift between even and odd components of the wave-function, thereby realizing an entangling $\sqrt{\text{SWAP}}$ gate.

The dynamics in the central harmonic trap are governed by the time-dependent Schr\"odinger equation for two particles \cite{busch1998two}: 
\begin{equation}\label{Eq_schroedinger_1}
\begin{split}
i\hbar \frac{\partial {{\psi }_{ss' }}}{\partial \tau}&=\big[
-\frac{{{\hbar }^{2}}}{2m}\left( \nabla _{x_1}^{2}+\nabla _{x_2}^{2} \right)+
V_{\rm ext}(x_1)+V_{\rm ext}(x_2)+\\
&+(1-\delta_s^{s'})\gamma\delta_{\text{reg}} (x_1-x_2) \big] {{\psi }_{ss'}} .
\end{split}
\end{equation}
where $V_{\rm ext}(x) = \frac12 m\omega_0^2x^2$ is the harmonic potential, and $\gamma$ is the interaction coupling constant governed by a magnetic field near the Feshbach resonance \cite{Chin2010}.
Changing to center-of-mass, $X=(x_1+x_2)/{\sqrt2}$, and relative, $x=(x_1-x_2)/{\sqrt2}$, coordinates decouples the problem: $\psi(X,x,\tau)=\psi_0(X,\tau)\psi_1(x,\tau)$. At $\tau_{\text{gate}}=\pi/\omega_0$, the $\psi_0$ part returns to its initial state, and the gate operation is determined entirely by $\psi_1(x,\tau)$.

In the non-interacting case ($\gamma=0$), the solutions for two wave packets separated by distance $d$ are displaced squeezed states,
\begin{equation}
\label{wave_packet_expr}
\varphi_\pm(x,\tau)=\Big(\tfrac{A(\tau)}{\sqrt{\pi}}\Big)^{1/2} e^{-i\Theta(\tau)/2} 
e^{-\frac{|x\pm x_c(\tau)|^2}{\sigma(\tau)^2}} e^{\pm i p_c(\tau)x/\hbar} ,
\end{equation}
with parameters
\begin{align}
A(\tau)&=\sqrt{\frac{m\omega_0 }{\hbar }}\frac{1}{\sqrt{\cosh (2r)+\sinh (2r)\cos (2\omega_0 \tau)}}\nonumber\\
\sigma(\tau)^2&=\frac{2\hbar }{m\omega_0 }\left[ \frac{1+\tanh (r){{e}^{-2i\omega_0 \tau}}}{1-\tanh (r){{e}^{-2i\omega_0 \tau}}} \right]\nonumber\\
{{e}^{-i\Theta (\tau)}}&={{\left[ \frac{1+\tanh (r){{e}^{+2i\omega_0 \tau}}}{1+\tanh (r){{e}^{-2i\omega_0 \tau}}} \right]}^{1/2}}{{e}^{-i\omega_0 \tau}}\nonumber\\
{{x}_{c}}(\tau)&=\frac{d}{\sqrt2}\cos (\omega_0 \tau)\nonumber\\
{{p}_{c}}(\tau)&=m{{\dot{x}}_{c}}(\tau)=-\frac{m\omega_0 d}{\sqrt2}\sin (\omega_0 \tau) \ \ .\label{wave_packet_expr_param}
\end{align}
The squeezing parameter, $r$, is determined by the initial condition $\sigma(0)=\sigma_0$,
\begin{equation}\label{Eq_squeezing_parameter}
r=\ln\left[\frac{\sigma_0}{\sqrt{2}\sqrt{\hbar/m\omega_0}}\right] \ \ .
\end{equation}

When interactions are included, the colliding wave-packets scatter with the delta potential at $x=0$ and the solution can be expressed as
\begin{equation}
\psi_1(x,\tau)=
\begin{cases}
\varphi_{+}(x,\tau)+R\varphi_{-}(x,\tau), & x>0 \\
T\varphi_{+}(x,\tau), & x<0 ,
\end{cases}
\end{equation}
with the reflection and transmission coefficients, $R$ and $T$, fixed by continuity and boundary conditions. A $\sqrt{\text{SWAP}}$ operation requires $R/T=\pm i$, which sets condition on $\gamma$ and $\tanh(r)$.
Numerical simulations of Eq.~\ref{Eq_schroedinger_1} using the beam propagation method (BPM) \cite{thylen1983beam} showed that this gate sequence can lead to fidelities close to 1, in various experimental settings.

\section{Verification of the numerical simulation}\label{sec_bloch}
Before presenting our gate sequence in an optical lattice, we first analyze the gate protocols reported in Ref.~\cite{Bojovic2026}. This comparison serves both to benchmark our numerical simulations and to highlight the advantages of the present scheme. 

In Ref.~\cite{Bojovic2026}, Bojović \emph{et al.} demonstrated high-fidelity collisional gates with two fermionic $^6$Li atoms confined in an optical superlattice (with short lattice constant $a_x = 1.14~\mu m$). The experiment realizes isolated double-well potentials formed by superimposing long and short optical lattices, where each double well hosts two atoms. 
The relevant dynamics are captured by the two-site Fermi–Hubbard Hamiltonian,
\begin{align}
\hat{H}_{\mathrm{FH}} =& -t\sum_{\sigma}(\hat{c}^\dagger_{L,\sigma}\hat{c}_{R,\sigma}
+\text{h.c.})+U\sum_{i=L,R}\hat{n}_{i,\uparrow}\hat{n}_{i,\downarrow}\nonumber\\
&+\frac{\delta}{2}\sum_{\sigma}(\hat{n}_{R,\sigma}-\hat{n}_{L,\sigma})
+\Delta_B(\hat{n}_{R,\uparrow}-\hat{n}_{R,\downarrow}),
\end{align}
where $t$ is the tunneling matrix element, $U$ the on-site interaction, $\delta$ the spin-independent energy bias between wells, and $\Delta_B$ a spin-dependent offset. The main experimental knobs are the lattice depths (controlling $t$ and $U$), the phase between long and short lattices (setting $\delta$), and the magnetic field gradient (tuning $\Delta_B$). The system can be initialized at will in any of the four basis states 
$|{\uparrow},{\downarrow}\rangle$, $|{\downarrow},{\uparrow}\rangle$, $|{\uparrow\downarrow},0\rangle$, or $|0,{\uparrow\downarrow}\rangle$, 
by applying controlled magnetic-field gradients and adjustable energy tilts between the wells.

While the tight-binding description suffices to analyze the gate sequences proposed in Ref.~\cite{Bojovic2026}, our protocol involves time dependences and parameter regimes that cannot be captured reliably by a two-site model alone. We therefore simulate the continuum, time-dependent Schr\"odinger equation for one or two particles (Eq.~\ref{Eq_schroedinger_1}).

In the tight-binding regime, relevant for comparison to Ref.~\cite{Bojovic2026}, the continuum parameters $(V,\gamma)$ map onto the Hubbard parameters $(t,U)$ via Wannier-like localized orbitals $w_L(x)$ and $w_R(x)$ centered in the left and right wells of a symmetric double well:
\begin{equation}\label{eq:U_g}
U \;=\; \gamma \int \! dx\, |w_L(x)|^4, 
\end{equation}
\begin{equation}\label{eq:t_eval}
t \;=\; -\!\int \! dx\, w_L^{\ast}(x)\left(-\frac{\hbar^2}{2m}\nabla^2 + V(x)\right) w_R(x) \, \, .
\end{equation}
We obtain $w_L(x)$ and $w_R(x)$ by diagonalizing the single-particle Hamiltonian of an isolated double well and forming symmetric and antisymmetric combinations of the two lowest eigenstates, yielding a pair of maximally localized orbitals. Extracting $(t,U)$ from $(V,\gamma)$ ensures parameter consistency between the continuum simulation and any tight-binding reference, thereby enabling a controlled verification against the measurements in Ref.~\cite{Bojovic2026}.

\subsection{Single-particle tunneling}

We first benchmark our numerics against the single-atom tunneling data reported in Extended Data Fig. 1a of Ref.~\cite{Bojovic2026}. In the experiment, single atoms are loaded into isolated double wells and initially pinned to one side by applying an energy offset. When this offset is reduced, the atom remains localized provided the short-lattice depth, $V_S$, is large (deep wells). Lowering $V_S$ to a target value initiates Rabi-like tunnel oscillations, and the population in the left well is sampled at various hold times. The resulting trace is fitted to a sinusoid, from which an oscillation frequency is extracted. Throughout these measurements the long-lattice depth is fixed at
\begin{equation}
V_L = 36.5\,E_r^{\mathrm{long}}, \qquad
E_r^{\mathrm{long}}=\frac{h^2}{8 m (2a_x)^2}\;.
\end{equation}
The short-lattice depth is varied and throughout this work will be given in units of $E_r^{\mathrm{short}}=\frac{h^2}{8 m a_x^2}$.

In the simulation we bypass the state-preparation sequence and start directly with the initial state being $w_L(x)$, at the desired lattice depths $(V_S,V_L)$. We then evolve $\psi(x,\tau)$ in time using the single particle version of Eq.~\ref{Eq_schroedinger_1} and monitor the left-well population
\begin{equation}
P_L(\tau)=\int_{-\infty}^{0}\! |\psi(x,\tau)|^2\,dx .
\end{equation}
As in the experiment, $P_L(\tau)$ is fitted to a sinusoid,
$P_L(\tau)=A\cos(2\pi f\,\tau+\varphi)+C$, and we identify $2t/h$ as the tunneling frequency. 

\begin{figure}[!t]
    \centering
    \includegraphics[width=\columnwidth]{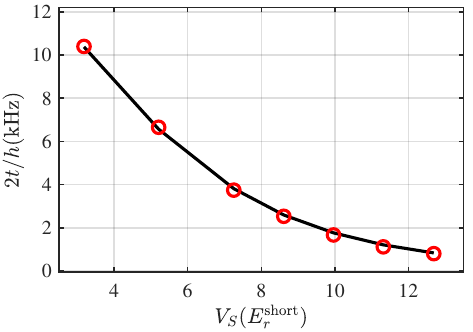}
    \caption{\textbf{Single-particle tunneling: simulation versus experiment.} Tunneling frequency as a function of the short-lattice depth $V_S$. Simulation (solid black) compared with the experimental data (red circles) from Extended Data Fig. 1a of Ref.~\cite{Bojovic2026}.}
    \label{fig:single}
\end{figure}

In Figure~\ref{fig:single} we present a comparison between the tunneling frequencies found in the experiment (red circles) and those calculated in our numerical simulations (black line). We find excellent quantitative agreement without parameter tuning: the maximal deviation is $\sim 8\%$, while the majority of points agree to within $2\%$. In addition, we calculated the tunneling frequency directly using Eq.~\ref{eq:t_eval}, and obtained the same values observed in the numerics. This confirms that the continuum simulation faithfully reproduces the single-particle tunneling dynamics under the experimental conditions.

\subsection{Two-particle tunneling}
As a second benchmark, we compare our results with the two-fermion measurements reported in Figure ~3 of Ref.~\cite{Bojovic2026}, which probe dynamics in the presence of repulsive interactions. Two experimental protocols are considered: (i) \emph{spin-exchange}, where two atoms with opposite spins are initialized on \emph{opposite} sites of a double well, and (ii) \emph{pair-tunneling}, where both atoms start on the \emph{same} site. The on-site interaction is tuned via a Feshbach resonance, and reported to be $U=h\times 6.7\,\mathrm{kHz}$. 
The long-lattice depth is held fixed at $V_L = 39.5\,E_r^{\mathrm{long}}$. The short-lattice depth is linearly ramped from $54\,E_r^{\mathrm{short}}$ down to $5.54\,E_r^{\mathrm{short}}$ over $500\,\mu\mathrm{s}$, held for a variable duration $\tau_h$, and then ramped back up over another $500\,\mu\mathrm{s}$. The relevant occupations, $P_{|\uparrow,\downarrow\rangle}$, for spin exchange, and $P_{|\uparrow\downarrow,0\rangle}$ for pair tunneling, are measured at the end of the full sequence.

In our simulations, we replicate the same ramp profiles and holding times, scanning $\tau_h\in[0,0.5]\,\mathrm{ms}$. Initial two-particle states are prepared either as opposite-spin fermions localized on different sites or co-localized on the same site. After time evolution, we project onto the site-occupation basis and record $P_{|\uparrow,\downarrow\rangle}$ and $P_{|\uparrow\downarrow,0\rangle}$ as functions of $\tau_h$, enabling a direct comparison with the experimental observables.

The key difference compared to the single-particle tunneling simulation is the inclusion of interparticle interactions. The contact-interaction coefficient $\gamma$ is obtained from Eq.~\ref{eq:U_g} for the double-well potential at the relevant double trap configurations, specifically $(V_L, V_S) = (34.9E_r^{\mathrm{long}}, 5.54E_r^{\mathrm{short}})$ of spin-exchange, and $(V_L, V_S) = (36.8E_r^{\mathrm{long}}, 6E_r^{\mathrm{short}})$  for pair tunneling. From the values of $U$ given in \cite{Bojovic2026} we deduce: $\gamma/h = 4.756\,\mathrm{kHz}\times\mathrm{\mu m}$ and  $\gamma/h = 4.361\,\mathrm{kHz}\times\mathrm{\mu m}$, for spin exchange and pair tunneling, respectively.

\begin{figure}[!t]
    \centering
    \includegraphics[width=\columnwidth]{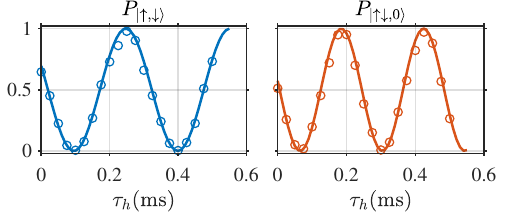}
    \caption{\textbf{Simulation of spin-exchange (left) and pair-tunneling (right) oscillations} for several holding times $\tau_h$. The fit to a sinusoidal form is shown as well as a solid line.}
    \label{fig:SE}
\end{figure}

We simulate both spin-exchange and pair-tunneling oscillations with these parameters, and results are presented in Figure~\ref{fig:SE}. Fitting the site-occupation traces to a sinusoidal form yields exchange frequencies $J/h = 3.324\,\mathrm{kHz}$ for spin-exchange, and $J/h = 4.182\,\mathrm{kHz}$ for pair tunneling. In the spin-exchange case, the agreement with the experimental results of Ref.~\cite{Bojovic2026} is excellent: our fitted frequency exactly matches the reported $3.32(3)\,\mathrm{kHz}$, and the extracted phase (i.e., the occupation at $\tau_h=0$) also coincides.

In the pair-tunneling regime, two discrepancies become apparent. First, the simulated oscillation frequency, $4.182\,\mathrm{kHz}$, exceeds the measured $3.8\,\mathrm{kHz}$ of Ref.~\cite{Bojovic2026} by a noticeable margin. Second, the fitted phase of the sinusoid differs from the experimental trace. Since these differences arise only in the pair-tunneling regime, we attribute them to deviations of the real interatomic interaction from the idealized delta-function potential, or from slight inaccuracies in the initial conditions.

\begin{figure*}[!t]
  \centering
  \includegraphics[width=\textwidth]{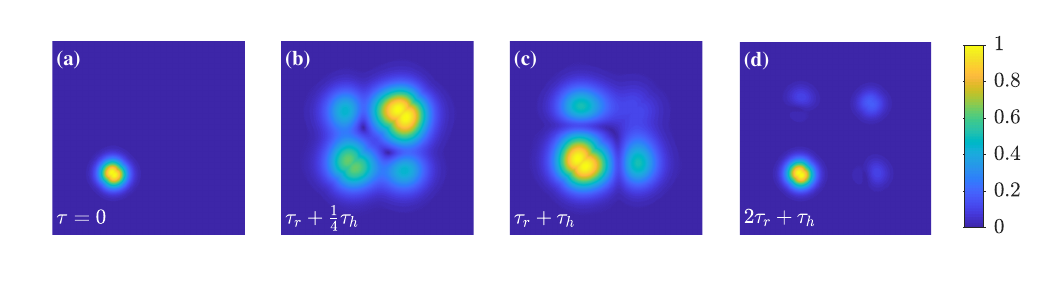}
  \caption{\textbf{Pair tunneling waveform evolution}. Panels (a)-(d) depict the absolute value of the two-particle waveform, $|\psi_{\uparrow \downarrow}(x_1,x_2,\tau)$, for different times during the operation of the sequence. $\tau_r = 0.5\,\mathrm{ms}$ - the ramping time, and $\tau_h = 0.2\,\mathrm{ms}$ - the holding time. $x_1$ and $x_2$ are the horizontal and vertical axes, respectively (each is $3\mathrm{\mu m}$ wide).}
  \label{fig:DH_2d_plot}
\end{figure*}

Figure \ref{fig:DH_2d_plot} illustrates the time evolution of the absolute value of the pair-tunneling wavefunction, $|\psi_{\uparrow \downarrow}(x_1, x_2, t)|$, at several representative moments. Initially, the two-particle wavefunction is localized in the left well. Owing to the repulsive delta-function interaction, the amplitude is slightly distorted along the diagonal $x_1 = x_2$ [Figure ~\ref{fig:DH_2d_plot}(a)]. As the barrier height is reduced, the pair begins to tunnel back and forth between wells, while part of the wavefunction separates, placing the two particles in opposite traps ($|x_1 - x_2| > d/2$) [Figure ~\ref{fig:DH_2d_plot}(b–c)]. When the barrier is raised again, the wavefunction becomes more confined at each site, and the probability distribution of particle positions depends sensitively on the holding time $\tau_h$ [Figure ~\ref{fig:DH_2d_plot}(d)].

\begin{figure}[!t]
    \centering
    \includegraphics[width=\columnwidth]{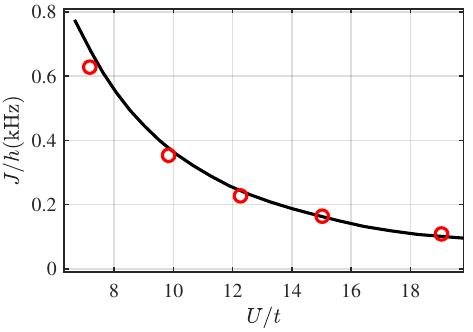}
    \caption{\textbf{Spin-exchange oscillations -- simulation versus experiment.}
The tunneling frequency is extracted from a series of simulations in which the short-lattice depth $V_S$ is varied while the interaction strength $\gamma$ is kept fixed. The on-site interaction $U$ is computed from Eq.~\ref{eq:U_g}; however, to quantitatively match the experimental data, it must be rescaled by a constant factor of $0.82$. This rescaling yields the solid black curve, which shows excellent agreement with the experimental results (red circles) reported in Extended Data Fig. 1b of Ref.~\cite{Bojovic2026} and follows the expected scaling $J = 4t^2/U$.}.
    \label{fig:tunneling_J}
\end{figure}

In addition, we benchmarked the spin-exchange oscillations in the strongly interacting regime $U/t \gg 1$, where the exchange frequency is well approximated by $J \simeq 4t^2/U$. 
Figure~\ref{fig:tunneling_J} compares our simulations with the experimental data (Extended Data Fig. 1b in \cite{Bojovic2026}), and shows very good agreement across the scanned parameters. Achieving quantitative consistency, however, requires rescaling the contact-interaction coefficient $\gamma$ by a constant factor of $0.82$.

\subsection{Tunneling--based $\sqrt{\text{SWAP}}$ gate sequences}

A key result of Ref.~\cite{Bojovic2026} is the experimental implementation of a tunneling-based $\sqrt{\text{SWAP}}$ gate. After establishing spin-exchange interactions, the authors proceed to realize two-qubit gates using controlled tunneling within the Fermi-Hubbard framework. The initial state consists of two atoms prepared in opposite wells, $\ket{\uparrow,\downarrow}$, separated by a high barrier ($V_S = 54E_r^\mathrm{short}$). To suppress unwanted pair-occupancy states ($\ket{\uparrow\downarrow,0}$ or $\ket{0,\uparrow\downarrow}$), they introduce three strategies:
(i) operate in the strong-interaction regime $U/t \gg 1$, where $J \approx 4t^2/U$, yielding very slow gates;
(ii) use fast ramps to increase and decrease the tunneling rate $t$, and fine-tune $U/t$ to $4/\!\sqrt{3}$, at which point the spin and charge sectors become independent at times $N \times h/(4J)$ (with $N$ an integer);
(iii) vary $t$ slowly compared to $U$, but still faster than $J$, or employ smoothly shaped Blackman pulses to obtain intermediate gate speeds. Using the last approach, the authors report a fidelity of $99.75\%$ for the Blackman-pulse sequence with a duration of 1.125 ms.

We follow the experimental protocols and, for strategy~(ii), use the reported ramp duration of $\tau_r = 50\,\mu\mathrm{s}$ while setting $V_L$ and $V_S$ according to Extended Data Table 1 of Ref.~\cite{Bojovic2026}. For an ideal $\sqrt{\text{SWAP}}$ and starting with $w_R(x1)w_L(x2)$, the resulting wave-function should be (up to a global phase factor) $\frac{1}{\sqrt{2}}\left[w_L(x_1)w_R(x_2)+
i w_R(x_1)w_L(x_2)\right]$. Therefore, we optimize the hold time, $\tau_h$, and the interaction strength, $\gamma$, to maximize the target-state fidelity:
\begin{equation}\label{process_fidelity}
\begin{split}
\mathcal{F}=& \Big|\big\langle\psi(x_1,x_2,\tau_{\rm gate})\big|\\ &\frac{1}{\sqrt{2}}\left[w_L(x_1)w_R(x_2)+
i w_R(x_1)w_L(x_2)\right]\big\rangle\Big|^2 ,
\end{split}
\end{equation}
where $\psi(x_1,x_2,\tau_{\rm gate})$ is the numerically calculated wavefunction at the gate end time $\tau_{\rm gate}$.

\begin{figure*}[!t]
  \centering
  \includegraphics[width=\textwidth]{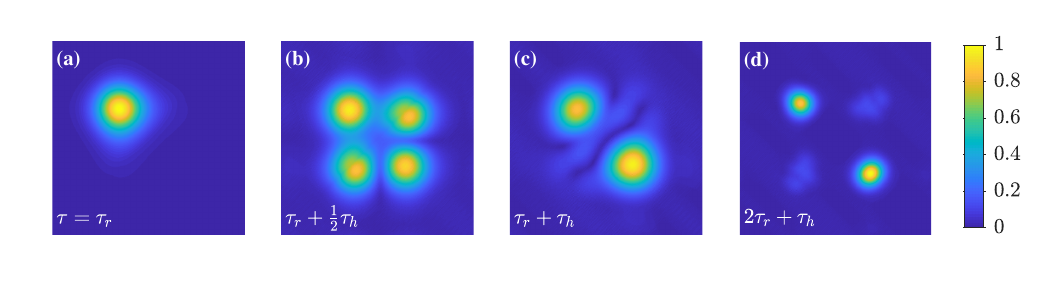}
  \caption{\textbf{$\sqrt{\mathrm{SWAP}}$ gate with fast ramping of amplitudes}. Panels (a)-(d) depict the absolute value of the two-particle waveform distribution, $|\psi_{\uparrow \downarrow}(x_1,x_2,\tau)|$, for different times during the operation of the sequence, with $x_1$ and $x_2$ being the horizontal and vertical axes, respectively (each is $3\mathrm{\mu m}$ wide).}
  \label{fig:fast_ramp}
\end{figure*}

We obtain that the optimized hold time and interaction coefficient are $\tau_h = 78.3\,\mu\mathrm{s}$ and $\gamma/h = 3.56\,\mathrm{kHz}\,\mu\mathrm{m}$, respectively. These values are close to the reported experimental parameters - $74.7\,\mu\mathrm{s}$ and $3.98\,\mathrm{kHz}\,\mu\mathrm{m}$~\cite{Bojovic2026}.
The highest fidelity reached using these parameters is $92.53\%$. The corresponding state populations are $P_{\ket{\uparrow,\downarrow}} = 0.5614$ and $P_{\ket{\downarrow,\uparrow}} = 0.3932$, while the doublon population indeed remains below $5\%$, in agreement with the observations of the experiment. The corresponding waveform evolution, shown in Figure~\ref{fig:fast_ramp}, exhibits sizable excursions outside the lowest-band manifold during the ramp. This leakage explains why a minimal two-site Fermi--Hubbard description cannot reproduce the full dynamics of the fast-ramp protocol.

For strategy~(iii), which uses a Blackman 
pulse, we similarly optimize both the pulse duration $\tau_p$ and the interaction parameter $\gamma$. We find that the optimal pulse duration is $\tau_p=1.0\,\mathrm{ms}$,  very close to the experimentally reported value of $1.125\,\mathrm{ms}$. The optimized interaction strength is found to be 
$\gamma/h = 3.0\,\mathrm{kHz}\,\mu\mathrm{m}$.

\begin{figure*}[!t]
  \centering
  \includegraphics[width=\textwidth]{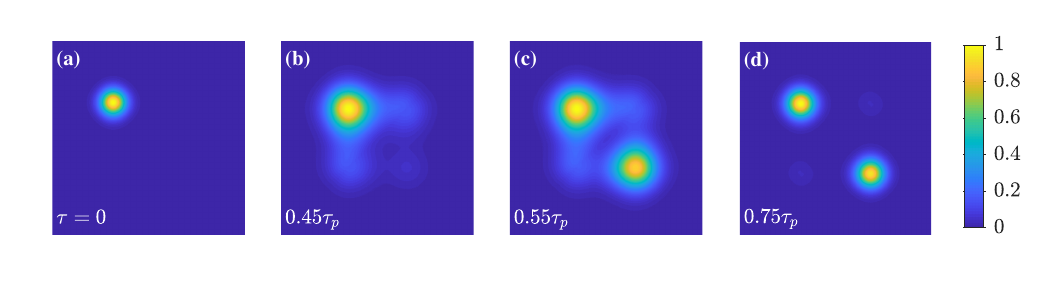}
    \caption{\textbf{$\sqrt{\mathrm{SWAP}}$ gate via Blackman pulse}. Panels (a)-(d) depict the absolute value of the two-particle wavefunction, $|\psi_{\uparrow \downarrow}(x_1,x_2,\tau)|$, for different times during the operation of the sequence, with $x_1$ and $x_2$ being the horizontal and vertical axes, respectively (each is $3\mathrm{\mu m}$ wide).}
    \label{fig:Blackman}
\end{figure*}

The highest fidelity we numerically get in the Blackman-pulse protocol is $99.78\%$, in excellent agreement with the experimental value of $99.75(6)\%$. The populations are $P_{\ket{\uparrow,\downarrow}} = 0.5303$ and $P_{\ket{\downarrow,\uparrow}} = 0.4685$. The corresponding state evolution, shown in Figure~\ref{fig:Blackman}, is nearly adiabatic: the wavefunction remains confined to the desired low-energy subspace throughout the sequence, consistent with the smooth spectral characteristics of the Blackman envelope.

\section{Fast SWAP using dynamic control of the superlattice}\label{sec_swap}
We now proceed to discuss our protocol for a fast $\sqrt{\mathrm{SWAP}}$ gate based on optimized, time-dependent control of the superlattice amplitudes $V_L(\tau)$ and $V_S(\tau)$. As a first step, we optimize \emph{single-particle} transfer from the left to the right well. With the interaction tuned to zero (e.g., via the magnetic field), this procedure realizes an effective SWAP operation for two particles through independent transport. Once this is accomplished, a $\sqrt{\mathrm{SWAP}}$ gate can be implemented by introducing back the interaction between the particles during the SWAP process~\cite{Nemirovsky}.

To implement a SWAP gate, we numerically solve Eq.~\ref{Eq_schroedinger_1} in the single-particle sector (coordinate $x$), using the superlattice potential $V_{\mathrm{ext}}(x,\tau)$ from Eq.~\ref{eq:super}. We restrict to experimentally convenient ranges $V_L<140\,E_r^{\mathrm{long}}$ and $V_S<50\,E_r^{\mathrm{short}}$. A naive strategy, namely rapidly turning off the short lattice (ramp $\sim 1\,\mu$s) while increasing the long lattice to its maximum, in the spirit of the added harmonic confinement in Sec.~\ref{sec_review}, performs poorly here. The reason is that  the residual long-lattice is appreciably anharmonic, which distorts the wave packet and limits the best achievable fidelity to $\sim 92.4\%$ ($\sim 85.4\%$) for a single- (two-) particle SWAP.

\begin{figure}[!t]
  \centering
  \includegraphics[width=\columnwidth]{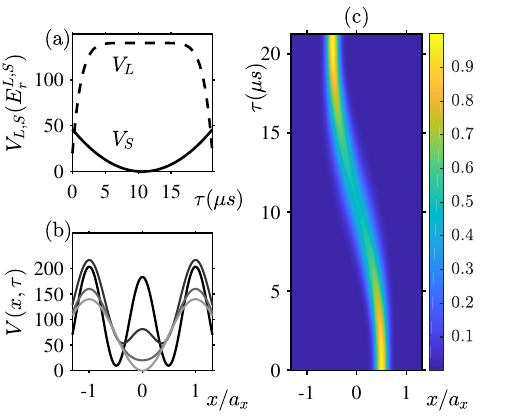}
  \caption{\textbf{Fast single-particle transfer for a SWAP.} 
  (a) Optimized envelopes for the short- and long-lattice depths, $V_S(\tau)$ and $V_L(\tau)$, over the gate duration $\tau_{\mathrm{gate}}$. 
  (b) Double-well potential snapshots every $\tau_{\mathrm{gate}}/6,$ from $\tau=0$ (black) to $\tau=0.5\,\tau_{\mathrm{gate}}$ (light gray). 
  (c) Single-particle probability density $|\psi(x,\tau)|^2$ (normalized to its maximum), illustrating the wave-packet transport from the left to the right well.}
  \label{fig:swap_1d}
\end{figure}

To mitigate the effect of anharmonicity, we instead ramp $V_L$ \emph{up} rapidly while lowering $V_S$ \emph{continuously} to (near) zero and subsequently restoring it, which preserves a quasi-harmonic confinement throughout the motion, as shown in Figure \ref{fig:swap_1d}. We use the following envelopes:
\begin{equation}\label{eq:Vs}
    V_S(\tau)= \frac{V_S(0)}{\tau_{\mathrm{gate}}^2}\,(2\tau-\tau_{\mathrm{gate}})^2
\end{equation}
\begin{equation}
V_L(\tau)= V_L^{\max}-\frac{V_L^{\max}-V_L(0)}{\tau_{\mathrm{gate}}^8}\,(2\tau-\tau_{\mathrm{gate}})^8 , 
\end{equation}
with $V_L^{\max}=140$, $V_L(0)=20$, $V_S(0)=45.93$ (in units of $E_r^{\mathrm{long}}$ and $E_r^{\mathrm{short}}$, respectively), and $\tau_{\mathrm{gate}}=21.26\,\mu\mathrm{s}$. The resulting total potential is depicted in Figure \ref{fig:swap_1d}b. 

With these parameters we achieve a transfer fidelity of $\sim 99.97\%$ for a single-particle and $\sim 99.94\%$ for the two-particle SWAP. Figure~\ref{fig:swap_1d}c depicts the corresponding dynamics: the wave packet travels swiftly from the left well to the central region and is subsequently recaptured in the right well, while remaining spectrally narrow and avoiding significant leakage to higher bands.

\section{Fast $\sqrt{\mathrm{SWAP}}$ gate}\label{sec_sqrt_swap}
Building on the single-particle transfer of Sec.~\ref{sec_swap}, implementing a $\sqrt{\mathrm{SWAP}}$ becomes straightforward: we solve the full two-particle Schr\"odinger equation (Eq.~\ref{Eq_schroedinger_1}) with the superlattice potential $V_{\mathrm{ext}}(x,\tau)$ of Eq.~\ref{eq:super}, driven by the optimized envelopes $V_L(\tau)$ and $V_S(\tau)$. The contact interaction is set by a single magnetic-field value $\gamma$ that is \emph{held constant} throughout the pulse. We then optimize over a small set of control parameters ($V_L(0)$, $V_S(0)$, $\tau_{\mathrm{gate}}$ and $\gamma$) to maximize the gate fidelity.

With this protocol we obtain the highest target-state fidelity (Eq. \ref{process_fidelity}) to be $99\%$ using the following parameters:
\begin{align}
    V_L^{\max}&=140E_r^{\mathrm{long}}\nonumber\\
    V_L(0)&=20E_r^{\mathrm{long}}\nonumber\\
    V_S(0)&=41.35E_r^{\mathrm{short}}\nonumber\\
    \tau_{\mathrm{gate}}&=21.2~\mu\mathrm{s}\nonumber\\
\gamma/h&=30.96~\mathrm{kHz}\times\mu\mathrm{m} \ \ .
\end{align}
Note that this quantity does not correspond to the gate fidelity, since it is evaluated for only a single initial state. In Sec.~\ref{sec:Gate Performance in the Computational Subspace}, we analyze the full average gate fidelity and obtain a value of $99.3\%$.

The duration of our proposed gate is shorter by a factor of $\sim 53$ relative to the experimental Blackman protocol, and by $\sim 8.4$ compared with a fast-ramp implementation (while still delivering much higher fidelity than the latter). As expected, the required interaction strength here is larger, by about a factor of six.

\begin{figure*}[!t]
  \centering
  \includegraphics[width=\textwidth]{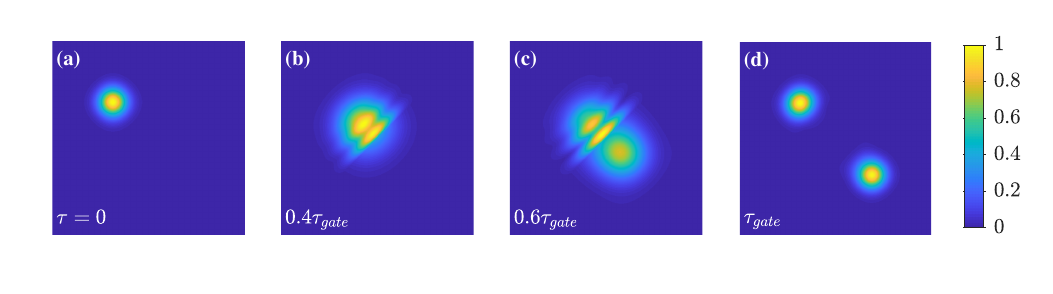}
    \caption{\textbf{$\sqrt{\mathrm{SWAP}}$ gate via fast dynamic control}. Panels (a)-(d) depict the absolute value of the two-particle wavefunction, $|\psi_{\uparrow \downarrow}(x_1,x_2,\tau)|$, for different times during the operation of the sequence, with $x_1$ and $x_2$ being the horizontal and vertical axes, respectively (each is $3\mathrm{\mu m}$ wide). $\tau_{gate} = 21.2\,\mu s$.}
    \label{fig:my_sqrt}
\end{figure*}

Figure~\ref{fig:my_sqrt} illustrates the two-particle dynamics during the pulse sequence. The joint wave packet moves toward the central region, where interaction-induced phase accumulation and exchange generate entanglement; the wave-packet components then separate and are recaptured in distinct wells at the end of the sequence. It is instructive to compare these dynamics with those of the Blackman gate shown in Figure~\ref{fig:Blackman}. A key qualitative difference is the appearance of cross-diagonal stripe patterns in the former, which are absent in the latter. These stripes arise from the interference between incoming and reflected wave packets and constitute a clear signature of a collisional process \cite{Nemirovsky}, whereas the dynamics in Figure~\ref{fig:Blackman} is dominated by tunneling.

Further improvement to the fidelity is possible if a larger superlattice depth is available. For example, increasing the ceiling to $V_L^{\max}=700 E_r^{\mathrm{long}}$ allows an optimized fidelity of $99.41\%$ with a shorter gate time $\tau_{\mathrm{gate}}=9.2~\mu\mathrm{s}$. This shorter gate requires a stronger contact interaction, $\gamma/h=63~\mathrm{kHz}\times\mu\mathrm{m}$.

These trends are consistent with the intuition from our previous work \cite{Nemirovsky}: higher effective confinement (i.e., stronger mid-pulse focusing) enhances the fidelity. A convenient proxy is a pseudo-squeezing parameter,
\[
r \;=\; \ln\!\Big[\sigma_0\big/\sigma(\tfrac{1}{2}\tau_{\mathrm{gate}})\Big],
\]
defined by the ratio of the wave-packet width at the start ($\sigma_0$) and at mid-gate. We find $r=-0.688$ for $V_L^{\max}=140$ and $r=-0.629$ for $V_L^{\max}=700$, in line with the observed fidelity gains as the mid-sequence confinement is strengthened.

\section{Robustness to lattice-amplitude variations}\label{sec:Robustness to lattice-amplitude variations}
In typical superlattice experiments, spatial intensity inhomogeneities of the lattice beams produce site-to-site depth variations at the $\sim\!5\%$ level \cite{chalopin2025optical}. Even a uniform rescaling of the control envelopes, $V_{L/S}(\tau)\!\to\!(1\!\pm\!0.05)\,V_{L,S}(\tau)$, degrades performance because the fixed timing no longer matches the effective tunneling rates. In our simulations, the single-pulse $\sqrt{\mathrm{SWAP}}$ fidelity decreases from $99\%$ to $\approx 97.5\%$ under a $\pm 5\%$ global scale change.

To mitigate this sensitivity, and to slightly increase the overall fidelity, we propose implementing the gate $\mathrm{SWAP}^{3/4}$ and repeating it twice, thereby realizing the maximally entangling gate $\mathrm{SWAP}^{3/2}$ (which is equivalent to $\mathrm{SWAP}\circ \sqrt{\mathrm{SWAP}}$). The improved fidelity of $\mathrm{SWAP}^{3/2}$ relative to $\sqrt{\mathrm{SWAP}}$ originates from two main effects. First, the required contact interaction strength is substantially reduced (by approximately $60\%$), which directly suppresses interaction-induced errors. The reduced sensitivity to amplitude variations arises from the approximate time-reversal symmetry of the repeated sequence: errors accumulated during the first application are partially canceled during the second. As a result, the double-$\mathrm{SWAP}^{3/4}$ protocol yields higher fidelities than a single $\mathrm{SWAP}^{3/4}$ gate in the presence of amplitude fluctuations.

With these modifications, and after optimizing both the lattice amplitudes and interaction strength, we obtain a fidelity of $99.3\%$ for the $\mathrm{SWAP}^{3/4}$ gate and $99.17\%$ for the composite $\mathrm{SWAP}^{3/2}$ gate. Importantly, under lattice-amplitude variations of up to $\pm 5\%$, the fidelity remains above $98.7\%$. A representative robust operating point is given by

\begin{align}
V_L^{\max}&=140E_r^{\mathrm{long}}\nonumber\\
V_L(0)&=20E_r^{\mathrm{long}}\nonumber\\
V_S(0)&=43.4E_r^{\mathrm{short}}\nonumber\\   
\tau_{\mathrm{gate}}&=2\times 21.39~\mu\mathrm{s}\nonumber\\
\gamma/h&= 12.8~\mathrm{kHz}\times\mu\mathrm{m}\, \, .
\end{align}

\section{Gate Performance in the Computational Subspace}\label{sec:Gate Performance in the Computational Subspace}

A convenient way to define the effective two-qubit operation and quantify the gate performance is to project the full time-evolution operator $U(\tau_g)$ onto the logical computational subspace spanned by
\[
\{|\uparrow\uparrow\rangle,\;|\uparrow\downarrow\rangle,\;|\downarrow\uparrow\rangle,\;|\downarrow\downarrow\rangle\}.
\]
This yields the projected operator
\[
M = P_{\mathrm{comp}}\, U(\tau_g)\, P_{\mathrm{comp}},
\]
where $P_{\mathrm{comp}}$ is the projector onto the computational subspace. In general, $M$ is not unitary, as it incorporates both coherent evolution within the logical subspace and leakage to states outside it. Its matrix elements are obtained from overlap integrals between the time-evolved wavefunctions and the initial logical states~\cite{jensen2019time}.

In the regime of one particle per well, it is convenient to express the dynamics in terms of symmetric and antisymmetric spatial wavefunctions constructed from localized orbitals $w_L(x)$ and $w_R(x)$,
\[
\psi_{\pm}(x_1,x_2)
=
\frac{1}{\sqrt{2}}\left[
w_R(x_1)w_L(x_2) \pm w_R(x_2)w_L(x_1)
\right],
\]
which correspond to the triplet ($+$) and singlet ($-$) sectors, respectively. Since the Hamiltonian is spin-independent, the dynamics in these sectors decouple, and the evolution is fully characterized by the return amplitudes
\[
a_{\pm} = \langle \psi_{\pm}(0)\,|\,\psi_{\pm}(\tau_g)\rangle.
\]

These amplitudes directly determine the projected operator $M$. In the singlet--triplet basis, the evolution is diagonal,
\[
M = a_{+} P_T + a_{-} P_S,
\]
where $P_T$ and $P_S$ are the projectors onto the triplet and singlet subspaces, respectively. Transforming back to the computational basis
$\{|\uparrow\uparrow\rangle,|\uparrow\downarrow\rangle,|\downarrow\uparrow\rangle,|\downarrow\downarrow\rangle\}$, one obtains
\[
M =
\begin{pmatrix}
a_{+} & 0 & 0 & 0 \\
0 & \tfrac{a_{+}+a_{-}}{2} & \tfrac{a_{+}-a_{-}}{2} & 0 \\
0 & \tfrac{a_{+}-a_{-}}{2} & \tfrac{a_{+}+a_{-}}{2} & 0 \\
0 & 0 & 0 & a_{+}
\end{pmatrix}.
\]

Using this procedure, we numerically evaluate the evolution starting from the states $\psi_{\pm}$. After removing an overall global phase, we obtain
\[
M =
\begin{pmatrix}
0.9962 & 0 & 0 & 0 \\
0 & 0.4960 + 0.4969 i & 0.5002 - 0.4969 i & 0 \\
0 & 0.5002 - 0.4969 i & 0.4960 + 0.4969 i & 0 \\
0 & 0 & 0 & 0.9962
\end{pmatrix}.
\]
The deviation of $M$ from unitarity reflects residual leakage from the computational subspace, which in the present case is dominated by excitation to higher vibrational modes, while population in the doublon-hole (DH) sector remains negligible (less than $2.7\cdot 10^{-4}$).

For the non-unitary projected operation $M$, we use the standard average gate fidelity formula for a quantum operation represented, within the computational subspace, by a single Kraus operator
\cite{Nielsen2002}. With respect to the ideal $\sqrt{\mathrm{SWAP}}$ unitary $U_{\mathrm{id}}$, it reads
\begin{equation}
\mathcal{F}_{\mathrm{avg}} =
\frac{\left|\operatorname{Tr}\!\left(U_{\mathrm{id}}^\dagger M\right)\right|^2 + d}{d(d+1)},
\qquad d=4,
\end{equation}
yielding $\mathcal{F}_{\mathrm{avg}} = 0.993$.

\section{Repeated Gate Performance}\label{sec:Repeated Gate Performance}

\begin{figure}[t]
\centering
\includegraphics[width=\columnwidth]{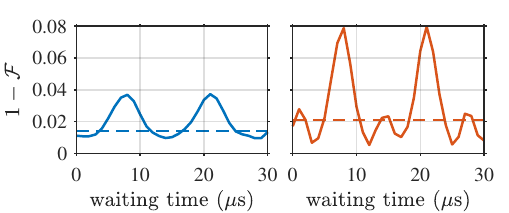}
\caption{\textbf{Repeated application of the $\sqrt{\mathrm{SWAP}}$ gate.}
Gate infidelity $(1-\mathcal{F})$ for two (left panel) and three (right panel) successive $\sqrt{\mathrm{SWAP}}$ operations, plotted as a function of the waiting time between consecutive gates.
The dashed lines indicate the naive expectation based on independent gate errors, assuming a single target-state fidelity of $\mathcal{F}=0.993$.}
\label{fig:repeat}
\end{figure}

We next investigate the performance of the gate under repeated application, focusing in particular on the effect of introducing a waiting time between successive gate operations.

We first apply the fast $\sqrt{\mathrm{SWAP}}$ sequence twice and compare the resulting evolution with the ideal $\mathrm{SWAP}$ operation. We then apply the sequence three times, with two equal waiting intervals inserted between consecutive gates, and compare the resulting operation with the ideal $\mathrm{SWAP}^{3/2}$ gate.

Figure~\ref{fig:repeat} shows the corresponding gate infidelity as a function of the waiting time between gate applications. The results demonstrate that phases accumulated in higher orbital modes, as well as in the doublon--hole (DH) sector, can significantly affect the performance of repeated gate sequences. Remarkably, interference between these contributions can also partially suppress the unwanted excitations, leading to fidelities that exceed the naive estimate $\mathcal{F}^n$, where $n$ is the number of applied gates (dashed lines in Figure~\ref{fig:repeat}).

Experimentally, a convenient method for estimating the gate fidelity is to repeatedly apply the gate sequence and measure the spin occupations after every four gates, for which the atomic state should ideally return to its initial configuration~\cite{Bojovic2026}.

\begin{figure}[t]
\centering
\includegraphics[width=\columnwidth]{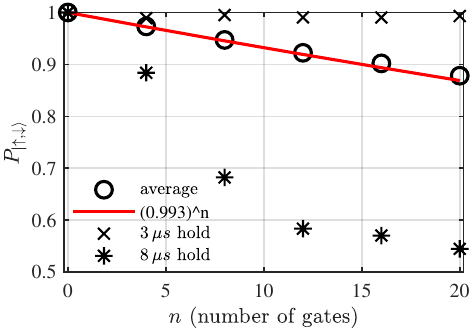}
\caption{\textbf{Performance of repeated gate sequences.}
Average return probability to the initial state after repeated applications of the $\sqrt{\mathrm{SWAP}}$ gate.
The statistics are obtained from 100 numerical realizations with randomized waiting times between consecutive gate operations.
Also shown are the best and worst realizations, corresponding to fixed waiting times.
The dashed curve shows the exponential decay model $(0.993)^n$, corresponding to independent errors per gate.}
\label{fig:repeat_probs}
\end{figure}

In our numerical simulations, we investigate the sensitivity of this procedure to the choice of waiting times between consecutive gates. To this end, the waiting times are randomly sampled from a uniform distribution in the range $[0,\,15\,\mu\mathrm{s}]$. Each realization consists of a sequence of 20 gate operations and 19 random waiting times. We generate 100 such realizations and evaluate the corresponding average return probability. The results are shown in Figure~\ref{fig:repeat_probs}. We also generate two additional trajectories with constant waiting times (chosen based on the results of Figure \ref{fig:repeat}), yielding the best and worst overall fidelity.

As can be seen, when the waiting times are randomized (circles), the decay follows an exponential behavior consistent with the expected single-gate infidelity (solid line). In contrast, for fixed waiting times, the decay curves can differ substantially, potentially leading to significant overestimation or underestimation of the fidelity of an individual gate. While some degree of randomization often arises naturally from experimental imperfections, these results demonstrate that this commonly used benchmarking procedure can be highly sensitive to coherent phase accumulation during the waiting intervals.

\section{Discussion}\label{sec_discussion}
We have introduced an optimized sequence for implementing a $\sqrt{\text{SWAP}}$ gate on a superlattice platform. This approach extends our previous work \cite{Nemirovsky}, in which two atoms were driven to interact within a central, ideal harmonic trap. In contrast, the present method relies solely on experimentally accessible control parameters of a superlattice. Since it does not require an additional external potential, it is much easier to implement in existing experiments.

Our results demonstrate that carefully timed modulation of the superlattice amplitudes enables coherent swapping of atoms between opposite lattice sites. When interatomic interactions are included, this mechanism naturally realizes a $\sqrt{\text{SWAP}}$ gate. The performance of the proposed protocol was investigated numerically using a model that goes beyond the tight-binding Fermi–Hubbard approximation. To validate our simulation framework, we benchmarked it against the experimental results reported in \cite{Bojovic2026}, finding excellent agreement, although our simulations were performed in one dimension. This correspondence strongly indicates that the proposed sequence can be implemented in realistic three-dimensional superlattices.

Finally, our numerical analysis shows that the gate exhibits high robustness, achieving fidelities exceeding 99\%. A key advantage of this protocol is its remarkably short operation time, on the order of $20\mu\text{s}$, which is roughly two orders of magnitude faster than tunneling-based gates (typically $\sim1\text{ms}$) \cite{Bojovic2026}.

\begin{acknowledgments}
This research was supported by the Pazy Research Foundation and by the Helen Diller Quantum Center. The data that support the findings of this article are openly available \cite{Data_Availability}.
\end{acknowledgments}

\textbf{Note added.} While finalizing this manuscript, we became aware of a closely related recent work~\cite{Reuter2025}.

%

\end{document}